\documentclass[letterpaper,prd,10pt,showpacs,showkeys,preprintnumbers,superscriptaddress,nofootinbib,preprint]{revtex4-1}
\usepackage{amsmath,amssymb,amsfonts,dsfont,mathrsfs,amsthm,mathtools,graphicx}
\usepackage{color}
\usepackage[usenames]{xcolor}
\usepackage{hyperref}
\usepackage{siunitx}
\hypersetup{linktocpage,colorlinks=true,urlcolor=blue,linkcolor=blue,citecolor=blue}  
\usepackage{bigints}
\usepackage[english]{babel}
\usepackage{float}

\def\rd{{\rm d}}

\newcommand*{\diag}{\operatorname{diag}}
\newcommand*{\Tr}{\operatorname{Tr}}
\newcommand*{\arctanh}{\operatorname{arctanh}}
\newcommand*{\arcsinh}{\operatorname{arcsinh}}

\newcommand{\diff}[1]{\text{d}#1}

\begin{document}

\title{Self-gravitating solutions in Yang-Mills-Chern-Simons theory coupled to 3D massive gravity}

\author{Crist\'obal \surname{Corral}}
\email{cristobal.corral@uai.cl}
\affiliation{Departamento de Ciencias, Facultad de Artes Liberales, Universidad Adolfo Ib\'a\~nez, Avenida Padre Hurtado 750, 2562340, Vi\~na del Mar, Chile}

\author{Daniel \surname{Flores-Alfonso}}
\email[]{danflores@unap.cl}
\affiliation{Instituto de Ciencias Exactas y Naturales, Universidad Arturo Prat, Avenida Playa Brava 3256, 1111346, Iquique, Chile}
\affiliation{Facultad de Ciencias, Universidad Arturo Prat, Avenida Arturo Prat Chac\'on 2120, 1110939, Iquique, Chile}

\author{Gast\'on Giribet}
\email{gaston.giribet@nyu.edu}
\affiliation{Department of Physics, New York University, 726 Broadway, New York, NY10003, USA.}

\author{Julio Oliva}
\email{juoliva@udec.cl}
\affiliation{Departamento de F\'isica, Universidad de Concepci\'on, Casilla, 160-C, Concepci\'on, Chile.}

\begin{abstract}
We study self-gravitating solutions of 3-dimensional massive gravity coupled to the Yang-Mills-Chern-Simons gauge theory. Among these, there is a family of asymptotically Warped-Anti de Sitter black holes that come to generalize previous solutions found in the literature and studied in the context of WAdS$_3$/CFT$_2$. We also present self-gravitating solutions to 3-dimensional Einstein-Yang-Mills theory, as well other self-gravitating solutions in the presence of higher-curvature terms.
\end{abstract}

\maketitle

\section{Introduction}

Three-dimensional gravity provides a tractable scenario to explore many aspects of gravity that are of insurmountable technical complexity in four and higher dimensions. Well known examples of this are the microscopic derivation of the black hole entropy beyond supersymmetry \cite{Strominger:1997eq} and the explicit computation of the gravity partition function \cite{Maloney:2007ud}. At a classical level, the 3-dimensional models provide a useful playground to study the non-linear regime of gravity and gauge theories; in particular, in the context of black hole physics. This is mainly due to the existence of the Ba\~nados-Teitelboim-Zanelli (BTZ) solution \cite{Banados:1992wn}, which describes black holes with physical properties that mimic those of their higher-dimensional analogs but, at the same time, resulting much more tractable. Being locally equivalent to 3-dimensional Anti-de Sitter (AdS$_3$) space \cite{Banados:1992gq}, the BTZ geometry also appears in any other theory that admits an AdS$_3$ vacuum. This includes 3-dimensional massive gravity, higher-spin theories, string theory and many others. A particularly interesting model to work with is the parity-even theory of massive gravity introduced in reference \cite{Bergshoeff:2009hq}, which is commonly known as New Massive Gravity (NMG). This is a higher-curvature theory that propagates two massive, spin-2 local degrees of freedom and exhibits interesting properties, such as the existence of a rich variety of black holes, apart from the BTZ solution. This includes generalizations of the BTZ black holes with a weakened AdS$_3$ asymptotics \cite{Oliva:2009ip, Bergshoeff:2009aq}, black holes with anisotropic scale invariance at infinity \cite{Ayon-Beato:2009rgu}, black holes in de Sitter (dS) space \cite{Oliva:2009ip}, and black holes in Warped AdS$_3$ (WAdS$_3$) spaces \cite{Clement:2009gq}. All these solutions are of interest in the context of holography, as they were used to explore to what extent the AdS/CFT duality could be generalized beyond asymptotically AdS$_3$ spaces \cite{Giribet:2009qz, deBuyl:2013ega, Gonzalez:2011nz, Donnay:2015iia, Donnay:2015vrb, Chernicoff:2024lkj}. Here, we will focus on the WAdS$_3$ spaces.

The WAdS$_3$ spaces appear in connection to black hole physics: they emerge in the near-horizon region of Kerr black holes \cite{Bengtsson:2005zj, Bardeen:1999px} and are crucial ingredients in the formulation of the Kerr/CFT correspondence \cite{Guica:2008mu}. Besides, the WAdS$_3$ spaces are at the root of the investigation of an entirely new class of conformal field theories (CFT) known as Warped-CFTs (WCFT) \cite{Detournay:2012pc}; see also \cite{Donnay:2015iia, Donnay:2015vrb}. Consequently, any generalization of WAdS$_3$ solutions, e.g. by coupling gauge fields, is in principle of interest both for black hole physics and for holography. Here, we will show that, when coupled to non-Abelian gauge theory, NMG admits black holes that asymptote WAdS$_3$ spaces. These are similar to the WAdS$_3$ black holes studied in \cite{Moussa:2008sj} and \cite{Clement:2009gq}, but, in contrast to those, the ones we construct here support a non-Abelian gauge field configuration. In fact, we will show that the gauge field may suffice to support the WAdS$_3$ black holes even in absence of higher-curvature terms. The model we will consider consists of Yang-Mills--Chern-Simons (YM-CS) gauge theory coupled to NMG. This results in a higher-derivative theory of gravity with a non-linear source, and that is the reason why, in order to solve the problem, we need a good strategy.

Our strategy goes as follows: Firstly, we recall that AdS$_3$ space can be written as a sort of Hopf fibration over its lower-dimensional analog, AdS$_2$, with the fiber being $\mathbb{R}$. This way of writing AdS$_3$ turns out to be particularly convenient to construct the 3-dimensional WAdS$_3$ spaces. The latter are easily obtained by introducing a real parameter ($\alpha$) that deforms the fibration over AdS$_2$, warping the geometry. In this way, one obtains either a stretched or squashed versions of AdS$_3$, in a similar manner as how a squashed 3-sphere is obtained from the round ${S}^3$. Secondly, we can resort to the results of \cite{Anninos:2008fx}, where the authors proved that the WAdS$_3$ black hole solutions previously found in \cite{Moussa:2008sj} turn out to be discrete quotients of the spacelike stretched (empty) WAdS$_3$ space, in an analogous way as how the BTZ black hole happens to be a discrete quotient of Lorentzian AdS$_3$ \cite{Banados:1992gq}. This means that, if we were able to prove that WAdS$_3$ spaces appear as non-vacuum solutions to the non-Abelian theory coupled to 3-dimensional massive gravity, then we would be ipso facto proving that the theory also admits charged WAdS$_3$ black holes. 

This strategy sounds simple; however, when trying to carry this out, one encounters a difficulty: Although WAdS$_3$ spaces are solvable, they are not Einstein manifolds, i.e. they are not of constant curvature, and they are not even conformally flat. In addition, they are not compact and of negative curvature. All this implies that, while not intractable, the metric of WAdS$_3$ black holes is much more involved than that of stationary BTZ. Besides, the theory we are dealing with is of fourth order in the metric, and non-linear both in the gravitational and in the matter sector. Thus, in order to circumvent the intrinsic difficulty of the problem, we find it convenient to look for these solutions progressively: In section~\ref{sec:II}, we introduce the theory. In section~\ref{sec:III}, we look for YM-CS solution on the 3-sphere with non-trivial gauge field configuration for the $SU(2)$ gauge field. The idea is to gain intuition from the compact case to select the right ansatz. In order to find such solutions, we propose a particularly friendly meron-type ansatz \cite{deAlfaro:1976qet} for the gauge field, which is shown to suffice. Then, we appropriately introduce a real parameter and obtain a squashed 3-sphere solution in the theory. The newly added degree of anisotropy requires us to modify the gauge potential accordingly. In turn, this change requires us to solve higher-order polynomials in order to present explicit solutions \cite{Canfora:2023bug}. Nonetheless, turning off the Chern-Simons term circumvents that difficulty. Thus, we concentrate on that case to present analytic anisotropic solutions. Finally, in section~\ref{sec:IV}, after having gained some intuition about the squashing procedure, we perform the analytic continuation to the non-compact $SU(1,1)$ group, which amounts to work out the WAdS$_3$ case, which has negative curvature. In this way, we finally find the WAdS$_3$ black holes that we are interested in. It turns out that the meron-type ansatz results in a useful trick to find exact solutions of YM-CS theory coupled to gravity. We emphasize this point by presenting more new solutions of NMG coupled to YM-CS gauge theory in the Appendix. We reserve section~\ref{sec:V} for some final remarks.

\section{Massive Gravity coupled to gauge theory\label{sec:II}}

We will be concerned with a 3-dimensional theory that consists of NMG coupled to YM-CS matter fields. More precisely, the gravity action is given by the Einstein-Hilbert term with a cosmological constant, supplemented by the higher-curvature terms introduced in~\cite{Bergshoeff:2009hq}. The matter sector is given by a non-Abelian $SU(2)$ field governed by the YM action augmented with a CS term. This yields
\begin{align}\label{action}
    I =  \kappa\int_{\mathcal{M}} \rd ^3x\sqrt{|g|}\left[R-2\Lambda - \frac{1}{\mu^2}\left(R_{\mu\nu} R^{\mu\nu} - \frac{3}{8}R^2 \right)  +  \frac{1}{2\kappa e^2}\Tr(F_{\mu\nu}F^{\mu\nu})\right] + \frac{4\pi^2 k}{e^2}I_{\rm CS}  \,,
\end{align}
where $\kappa=(16\pi G)^{-1}$ is the gravitational coupling, $\Lambda$ is the cosmological constant, and $\mu$ is a dimension-1 parameter. We use the standard notation for this last coupling constant, $\mu^2$, which after linearization represents the graviton mass. In the matter sector, $e$ denotes the coupling constant of the gauge theory. The non-Abelian field strength is defined as $F_{\mu\nu}=\partial_\mu A_\nu - \partial_\nu A_\mu + [A_\mu,A_\nu]$, which, expressed in terms of differential forms, yields the curvature 2-form $F=\tfrac{1}{2}F_{\mu\nu}\diff{}x^{\mu}\wedge \diff{}x^{\nu}=\diff{A}+\tfrac{1}{2}[A,A]$. The gauge field decomposes as usual, $A=A_{\,\mu}\diff{}x^\mu =A^i_{\,\mu}t_i\diff{}x^\mu $, with $i=1,2,3$ being the group indices; $t_i$ are the generators of ${su}(2)$ algebra. When considering ${su}(1,1)$ algebras, we use $\eta_i$ for the generators instead. The action also includes a CS term 
\begin{align}
    \label{CSaction}
    I_{\rm CS} = \frac{1}{8\pi^2}\int_{\mathcal{M}} \Tr\left(A\wedge\diff{A} + \frac{2}{3}A\wedge A\wedge A \right)\,,
\end{align}    
which is natural in dimension 3. We will omit here the discussion about the topological origin of (\ref{CSaction}), the role of large gauge transformations, the quantization condition for $k$ for compact groups, and all that. Also, while the solutions we will discuss here admit a straightforward generalization to the case in which a gravitational CS term \cite{Deser:1981wh, Deser:1982vy} is also included, here we will not deal with it.

The field equations derived from the action above take the form
\begin{subequations}\label{eom}
    \begin{align}\label{eomg}
    \mathcal{E}^{(g)}_{\mu\nu} &\equiv R_{\mu\nu} - \frac{1}{2}g_{\mu\nu}R + \Lambda g_{\mu\nu}  - \frac{1}{2\mu^2}K_{\mu\nu} - \frac{1}{2\kappa}T_{\mu\nu} = 0\,, \\
    \mathcal{E}_{(A)}^\mu &\equiv \nabla_\nu F^{\nu\mu} + [A_\nu,F^{\nu\mu}] + k\,\varepsilon^{\mu\nu\lambda}F_{\nu\lambda} = 0\, ,
    \label{eomA}
\end{align}
\end{subequations}
where
\begin{align}\label{Kmunu}
    K_{\mu\nu} &= 2\Box R_{\mu\nu} - \frac{1}{2}\nabla_\mu\nabla_\nu R - \frac{1}{2}g_{\mu\nu}\Box R + 4R_{\mu\lambda\nu\rho}R^{\lambda\rho}-\frac{3}{2}RR_{\mu\nu} - g_{\mu\nu}R_{\lambda\rho}R^{\lambda\rho} + \frac{3}{8}g_{\mu\nu}R^2, \\
    \label{Tmunu}
    T_{\mu\nu} &= -\frac{2}{e^2}\Tr\left(F_{\mu\lambda}F_{\nu}{}^{\lambda} - \frac{1}{4}g_{\mu\nu}F_{\lambda\rho}F^{\lambda\rho} \right)\,.
\end{align}
Despite the field equations~\eqref{eomg} are of fourth order in the metric, their trace is of second order. This is at the very root of the interesting properties of NMG, e.g. the absence of ghosts about its maximally symmetric vacua.

This theory admits many interesting solutions in vacuum, including hairy black holes \cite{Oliva:2009ip, Bergshoeff:2009aq}, warped black holes~\cite{Clement:2009gq}, Lifshitz black holes~\cite{Ayon-Beato:2009rgu}, gravitational waves~\cite{Ayon-Beato:2009cgh}, and many other spacetimes; see~\cite{Flores-Alfonso:2020zif} and references therein. Solutions in the presence of gauge fields, in contrast, are much harder to obtain, at least analytically, cf.~\cite{Banados:2005da,Moussa:2008sj,Ghodsi:2010gk}. In fact, due to the higher-derivative terms of NMG, a seemingly simple problem such as charging a BTZ black hole with a $U(1)$ gauge field turns out to be rather difficult. Despite this difficulty, here we will be able to find solutions with non-Abelian gauge fields, enlarging in this way the set of known solutions of massive gravity in the presence of matter.

\section{Self-gravitating solutions with $SU(2)$ gauge field\label{sec:III}}

As explained in the introduction, our first step will be finding solutions of positive curvature with non-trivial $SU(2)$ field configuration. That is to say, we look for solutions to Eqs.~\eqref{eom} by considering an ${su}(2)$-valued gauge connection $A=A_\mu^i t_i\diff{}x^\mu$, where $i=1,2,3$ are indices of the group, $t_i=-\frac{i}{2}\tau_i$ are generators of the $SU(2)$ group satisfying the ${su}(2)$ Lie algebra $[t_i,t_j]=\epsilon_{ijk}t^k$, and $\tau_i$ are the Pauli matrices. Notice that we have chosen the generators to be anti-hermitian. The standard (anti)commutation relations of the latter yield $\Tr(t_it_j) = -\tfrac{1}{2}\delta_{ij}$ and $\Tr(t_it_jt_k) = - \frac{1}{4}\epsilon_{ijk}$. 

In the following, we construct different self-gravitating $SU(2)$ gauge configurations that support special 3-dimensional geometries in NMG.

\subsection{The round 3-sphere\label{sec:S3round}}

The simplest regular space with positive curvature in three dimensions is the 3-sphere, ${S}^3$. To obtain self-gravitating YM-CS configurations that support this geometry in NMG, it is convenient to introduce the left-invariant forms of $SU(2)$, call them $\sigma_i$, which can be parameterized in terms of the Euler angles,
\begin{subequations}
    \begin{align}
    \sigma_1 &= \cos\psi\,\diff{\vartheta} + \sin\vartheta\sin\psi\,\diff{\varphi}\,, \\
    \sigma_2 &= -\sin\psi\,\diff{\vartheta} + \sin\vartheta\cos\psi\,\diff{\varphi}\,, \\
    \sigma_3 &= \diff{\psi} + \cos\vartheta\,\diff{\varphi}\,,
\end{align}
\end{subequations}
where $0\leq\vartheta\leq\pi$, $0\leq\varphi<2\pi$, and $0\leq\psi\leq 4\pi$. These $1$-forms satisfy the Maurer-Cartan equations of $SU(2)$, namely $\diff{\sigma_i} + \tfrac{1}{2}\epsilon_{ijk}\sigma^j\wedge\sigma^k=0$. The metric of ${S}^3$ can be written in terms of the left-invariant forms of $SU(2)$ as  follows
\begin{align}\label{S3metric}
    \diff{s^2} = \rho_0^2\; \delta_{ij}\,\sigma^i \otimes \sigma^j = \rho_0^2\left[\left(\diff{\psi} + \cos\vartheta\diff{\varphi} \right)^2 + \diff{\vartheta^2} + \sin^2\vartheta\diff{\varphi^2} \right]\,,
\end{align}
where $\rho_0\in \mathbb{R}_{+}$ is a constant that denotes the radius of ${S}^3$. The metric is invariant under the action of the $SO(4)$ isometry group. For the non-Abelian field $A$, we assume an ansatz aligned along the left-invariant forms, i.e.
\begin{align}\label{Aansatz1}
    A = A_\mu^i t_i \diff{x^\mu} = \lambda\,\sigma^i\,t_i\,,
\end{align}
where $SU(2)$ indices are raised and lowered with the Cartan-Killing metric. Despite the remarkably simple ansatz for $A$, the non-linearity of the gauge theory yields a non-vanishing field strength, provided as $0 \neq \lambda \neq 1$; namely
\begin{align}\label{Fmeronic}
    F = \frac{\lambda}{2}(\lambda-1)\epsilon^{ijk}\,t_i\,\sigma_j\wedge\sigma_k\,.
\end{align}
In Euclidean 4-dimensional space, solutions of this sort were first studied in Ref.~\cite{deAlfaro:1976qet}, for which the YM equations demand $\lambda=1/2$. Such configurations were dubbed \emph{merons}, and present localized fractional topological charge; see also the early works ~\cite{Callan:1977gz,Callan:1977qs,Glimm:1977sx,Callan:1978bm,Actor:1979in}. Black hole solutions in dimension 4 with similar gauge field configurations were recently studied in~\cite{Canfora:2012ap}.

Inserting the ans\"atze for the metric and the gauge fields into the field equations~\eqref{eom}, we find that the latter are solved if
\begin{subequations}\label{S3sol}
    \begin{align}
    \lambda &= \frac{1}{2} + k\rho_0\,, \\
    \rho^2_{0} &=  \frac{1}{4\left(4\kappa\Lambda e^2-k^4\right)}\left\{2\kappa e^2 - k^2 \pm e\sqrt{\kappa\left[4\left(\Lambda+\kappa e^2 - k^2\right) - \frac{1}{\mu^2}\left(4\kappa\Lambda e^2-k^4 \right)\right]} \right\}\,,
\end{align}
\end{subequations}
for $4\kappa\Lambda e^2-k^4\neq0$. Notice that, in dimension 3, the presence of the CS term produces a shift in the value $\lambda=1/2$. This was first observed in the context of the 3-dimensional Einstein-YM-CS theory in Ref.~\cite{Canfora:2017yio}. In fact, our solution (\ref{S3sol}) reduces to those found in~\cite{Canfora:2017yio} in the limit  $\mu\to\infty$. Notice that, for $\rho_0^2$ to be positive, the parameters of the theory have to be constrained. Moreover, in the {topological limit} $e\to 0 $, $k\to 0$ with $|k/e|<\infty$, where the matter content reduces to CS, one recovers the pure gauge solutions $\lambda =0$ and $\lambda=1$. If the CS coupling $k$ vanishes, there exist a perturbative branch for which $\Lambda\to0$, i.e.
\begin{equation}
    \lambda = \frac{1}{2}\,, \ \ \ \ \ \ \ 
    \rho_0^2 = \frac{1}{16\mu^2} - \frac{1}{16\kappa e^2} + \mathcal{O}(\Lambda)\,.
\end{equation}

The particular case $k^4=4\kappa\Lambda e^2$ needs to be studied separately. In that case, the solution becomes
\begin{equation}
    \lambda = \frac{1}{2} + k\rho_0\,, \ \ \  \ \ 
    \rho_0^2 = \frac{\kappa e^2 - \mu^2}{8\mu^2\left(2\kappa e^2 - k^2 \right)}=\frac{k^4-4\mu^2 \Lambda }{16\mu^2 k^2(k^2-2\Lambda )}\,. \
\end{equation}
The gauge field configuration that supports this solution has a nontrivial topological charge, which can be seen by integrating the CS term~\eqref{CSaction} evaluated on the configuration~\eqref{Aansatz1}, giving
\begin{align}
    I_{\rm CS} = \frac{1}{2} + \frac{3k\rho_0}{2} - 2k^3\rho_0^3\,.
\end{align}

Now that we have shown that NMG coupled to YM-CS theory admits a 3-sphere with a non-trivial gauge field on it, we will introduce a squashing parameter and try to see how the solution can be gently deformed. 

\subsection{The squashed 3-sphere\label{sec:squashedS3}}

Deforming one of the directions of the 3-sphere, say $\sigma_3$, by adding a warping factor $\alpha \in \mathbb{R}_+$ leads to a space of non-constant curvature. The metric of such space is that of the squashed 3-sphere, which can be locally written as
\begin{equation}\label{ESquashed}
    \rd s^2 = \rho_0^2\left(\sigma_1^2 + \sigma^2_2 + \alpha^2\sigma_3^2 \right) = \rho_0^2\left[\rd\vartheta^2 + 
\sin^2\vartheta\,\rd\varphi^2+\alpha^2(\rd \psi+\cos\vartheta\,\rd \varphi)^2\right]\,,
\end{equation}
with $\rho_0$ being the radius of the squashed 3-sphere. This metric is no longer invariant under the action of the isometry group $SO(4)=SU(2)\times SU(2)\,/\mathbb{Z}_2$; its symmetry group is broken down to $SU(2)\times U(1)$ due to squashing; $\alpha$ can be seen as a symmetry-breaking parameter. The squashed 3-sphere is an $\eta$-Einstein space, which means that it is equipped with a contact 1-form, $\eta=\eta_\mu\diff{x^\mu}$, and it satisfies the condition~\cite{Okumura:1962,Blair:2002}
\begin{equation}
    R_{\mu\nu} =  a\,\eta_\mu\eta_\nu + b\, g_{\mu\nu} , \label{etaEinstein}
\end{equation}
for some $a$ and $b$. These geometries have been studied in Refs.~\cite{Flores-Alfonso:2020zif,Andrianopoli:2023dfm,Andrianopoli:2024twc}. In the case of the squashed 3-sphere, the contact form is $\eta=\sigma_3$ and the values of $a$ and $b$ are
\begin{equation}
    a=\alpha^2(\alpha^2-1), \ \ \ \ \ b=\frac{2-\alpha^2}{2\rho_0^2}.
\end{equation}
Since these constants only depend on $\alpha^2$ and the radius, one can perform the analytic continuation $\alpha\to i\alpha$ and obtain a squashed analog of the dS$_3$ space.

To obtain a squashed 3-sphere solution with non-trivial $SU(2)$ gauge field configuration in NMG, we take
\begin{align}
A = \sum_{i=1}^{3}\lambda_{(i)}\sigma_i t_i\, , \label{Aansatz2}
\end{align} 
as has been explored in Einstein gravity~\cite{Canfora:2023bug}. There, the solution for the YM-CS equations was found and remains unchanged in the NMG case. The relevant point here is that the matter equations fix the parameter $\lambda_{(i)}$ in terms of the squashing parameter and the CS coupling. The latter enters into the gravity sector only through $\lambda_{(i)}$ since, recall, the CS term is absent from the energy-momentum tensor due to its topological nature. 

Although analytic solutions to the YM-CS equations can be obtained for $\lambda_{(i)}$, the metric cannot be obtained in the same fashion as~\eqref{S3sol} due to the presence of fifth-order polynomials in $\rho_0$. However, for $k=0$, the result is simply
\begin{subequations} \label{lambdas}
\begin{align}
\lambda_{(1)}^2 &  = \lambda_{(2)}^2 =  \frac{\alpha}{8}\left(  \sqrt{9\alpha^{2}+16}%
-3\alpha\right)  \,,\\
\lambda_{(3)} &  =1-2\lambda_{(2)}^2\,,
\end{align}
\end{subequations}
which has nontrivial topological charge as shown in Ref.~\cite{Canfora:2023bug}. Inserting  Eq.~\eqref{ESquashed} and the ansatz~\eqref{Aansatz2} into the field equations~\eqref{eom}, we find that solutions to the NMG equations sourced by YM matter are given by
\begin{subequations}\label{SQsol}
    \begin{align}
\rho_0^2 &  = \frac{ (\lambda_{(2)}-\lambda_{(2)}\lambda_{(3)})^2-\alpha^2(\lambda_{(2)}^2-\lambda_{(3)})^2 }{2\kappa e^2\alpha^2(\alpha^2-1)} + \frac{4-21\alpha^2}{8\mu^2}  \,,\\
\Lambda &  =\frac{\alpha^2}{4\rho_0^2} + \frac{63\alpha^4-80\alpha^2+16}{64\mu^2\rho_0^4} + \frac{(\lambda_{(2)}^2-\lambda_{(3)})^2}{4\kappa e^2\rho_0^4} \,.
\end{align}
\end{subequations}

These results showcase how anisotropic meron type gauge fields are able to source gravitational backgrounds with a similar kind of anisotropy. Hence, we expect similar fields to exist for WAdS$_3$ spacetimes. 

\section{Warped AdS black hole\label{sec:IV}}

Having studied the case of the 3-sphere and the squashed 3-sphere, we are ready to present the solution for the WAdS$_3$ black hole. As explained in the introduction, the argument goes as follows: AdS$_3$ space can be written as a Hopf fibration over AdS$_2$, just like in the case of the 3-sphere, by simply replacing trigonometric functions in Eq.~\eqref{S3metric} by their hyperbolic counterparts. Then, the WAdS$_3$ spaces are nothing but the $\alpha$-deformations of AdS$_3$, exactly as done in (\ref{ESquashed}) for the squashed 3-sphere. Having solved the case of positive curvature, the hyperbolic analog can easily be obtained by continuing the gauge theory to the $SU(1,1)$ group; the latter is isomorphic to $SL(2,\mathbb{R})$. For further details on how this process is carried out see Sec. 4 of \cite{Duff:1998cr}.

We begin by writing warped AdS$_3$ space in analogy with \eqref{ESquashed}, i.e., 
\begin{equation}
    \diff{s^2} =  \rho_0^2\left[ - \cosh^2\sigma\diff{\tau^2} + \diff{\sigma^2} + \alpha^2\left(\diff{u} + \sinh\sigma\diff{\tau} \right)^2 \right]\,.
\end{equation}
At the same time, we require the left-invariant forms of the gauge group to construct a gauge potential that serves our purposes. In the hyperbolic coordinates used above, they can be parametrized as
\begin{subequations}
    \begin{align}
    \omega_1 &= -\sinh u\diff{\sigma} + \cosh u \cosh\sigma\diff{\tau}\,, \\
    \omega_2 &= \cosh u\diff{\sigma} - \sinh u \cosh\sigma\diff{\tau}\,, \\
    \omega_3 &= \diff{u} + \sinh\sigma\diff{\tau} \,.
\end{align}
\end{subequations}
These $1$-forms satisfy the Maurer-Cartan equations $\diff{\omega_i} + \tfrac{1}{2}f_{ijk}\omega^j\wedge\omega^k=0$, with group indices being raised and lowered with the Cartan-Killing metric of $SU(1,1)$, say $\eta_{kl} = \diag(-1,1,1)$, with $f_{ijk}=\epsilon_{ijl}\eta_{kl}$ the structure constants thereof. Moreover, in terms of them, the spacetime metric takes the form $\diff{s^2} = \rho_0^2\;\left( -\omega_1^2+\omega_2^2+\alpha^2\omega_3^2\right)$, bearing a close resemblance with that of Eq.~\eqref{S3metric}. For the non-Abelian gauge fields, we assume the following ansatz
\begin{align}
A = \sum_{i=1}^{3}\lambda_{(i)}\omega_i\eta_i\, ,\label{elgauga}
\end{align}
where the $\lambda_{(i)}$ are constants, and $\eta_i$ are the generators of the gauge group satisfying the ${su}(1,1)$ Lie algebra $[\eta_i,\eta_j]=f_{ijk}\eta_k$. 

Following the same steps as before, but adapting them to the hyperbolic case, we are led to find the adequate relations between the coupling constants of the theory and the squashing parameter $\alpha$. We also take $k=0$, as in the previous section, to avoid higher-degree polynomials. Then, the field equations of NMG coupled to Yang-Mills fields are solved by
\begin{subequations}\label{WAdSsol}
    \begin{align}
\rho_0^2 &  = \frac{ (\lambda_{(2)}-\lambda_{(2)}\lambda_{(3)})^2-\alpha^2(\lambda_{(2)}^2-\lambda_{(3)})^2 }{2\kappa e^2\alpha^2(\alpha^2-1)} - \frac{4-21\alpha^2}{8\mu^2}  \,,\\
\Lambda &  = -\frac{\alpha^2}{4\rho_0^2} + \frac{63\alpha^4-80\alpha^2+16}{64\mu^2\rho_0^4} - \frac{(\lambda_{(2)}^2-\lambda_{(3)})^2}{4\kappa e^2\rho_0^4} \,, \label{lalphaWAdS}
\end{align}
\end{subequations}
while the $\lambda_{(i)}$ parameters satisfy the same conditions as the squashed sphere, Eq.~\eqref{lambdas}. All the parameters get fixed in terms of $\alpha$ which, in turn, is a function of the cosmological constant. Notice that this solution can be obtained from Eq.~\eqref{SQsol} by performing $\mu^2\to -\mu^2$ and $\Lambda\to-\Lambda$. 

Now, we take into account the observation of Ref.~\cite{Anninos:2008fx}, which states that the WAdS$_3$ black holes of \cite{Moussa:2008sj} are discrete quotients of WAdS$_3$. Thus, we are ready to derive the solutions we are looking for. First, we perform the change of coordinates 
\begin{subequations}
\begin{align}\notag
    u &= \frac{\ell^2}{4\alpha\rho_0^2} \sqrt{\frac{4-\alpha^2}{3}}\left[2t - \left(\alpha\sqrt{\frac{3}{4-\alpha^2}}(r_++r_-) + \frac{\ell\sqrt{r_+r_-}}{\rho_0} \right)\theta \right] \\ &\qquad - \arctanh\left[\frac{r_++r_--2r}{r_+-r_-}\coth\left(\frac{\ell^2(r_+-r_-)}{4\rho_0^2}\,\theta\right) \right]\,, \\
    \sigma &= \arcsinh\left[\frac{2\sqrt{(r-r_+)(r-r_-)}}{r_+-r_-} \cosh\left(\frac{\ell^2(r_+-r_-)}{4\rho_0^2}\,\theta \right) \right]\,, \\
    \tau &= \arctan\left[\frac{2\sqrt{(r-r_+)(r-r_-)}}{2r-(r_++r_-)} \sinh\left(\frac{\ell^2(r_+-r_-)}{4\rho_0^2}\,\theta \right) \right]\,,
\end{align}    
\end{subequations}
which are defined for the range $1\leq\alpha<2$ (see~\cite{Anninos:2008fx}). Then, as it happens for NMG in vacuum~\cite{Clement:2009gq}, one finds that the field equations~\eqref{eom} are solved by the stationary metric
\begin{align}
    \diff{s^2} = -f(r)\,\diff{t^2} + \frac{\diff{r^2}}{h(r)} + \rho^2(r)\left(\diff{\theta} + N^\theta(r)\,\diff{t} \right)^2 \,,
\end{align}
with the metric functions
\begin{subequations}
    \begin{align}
    f(r) &= \frac{\ell^2(r-r_+)(r-r_-)}{r\left[2\alpha\sqrt{r_+r_-}+(\alpha^2-1)r + r_+ + r_- \right]} \,, \label{LAPRIMERA}\\
    h(r) &= \frac{(r-r_+)(r-r_-)}{\rho_0^2} \,, \\
    \rho^2(r) &= \frac{r\ell^4\left[2\alpha\sqrt{r_+r_-}+(\alpha^2-1)r + r_+ + r_- \right]}{4\rho_0^2}\,, \\
    N^\theta(r) &= \frac{2\rho_0\left[\sqrt{r_+r_-} + \alpha\,r \right]}{r\ell\left[2\alpha\sqrt{r_+r_-}+(\alpha^2-1)r + r_+ + r_- \right]}\,,
\end{align}
\end{subequations}
where $\Lambda=-\ell^{-2}$. The two integration constants $r_\pm$ represent the locations of the outer and the inner horizons of the black hole provided $0\leq r_- \leq r_+$, cf. \cite{Anninos:2008fx}.

The field configuration~\eqref{elgauga} can be seen to be finite at the horizon and to exhibit an asymptotic behavior similar to the one obtained in \cite{Banados:2005da} for the $U(1)$ gauge field in asymptotically WAdS$_3$ spaces. Additionally, the conserved charges associated with the WAdS$_3$ black holes have been computed in \cite{Donnay:2015iia}; see also references therein and thereof. A similar computation works here, as the asymptotic behavior of the Yang-Mills fields translates into a shift of $\rho_0$ and the WAdS$_3$ curvature radius of the black hole. Then, in our notation, the mass and angular momentum of the WAdS$_3$ black hole are~\cite{Donnay:2015iia}
\begin{align}
    Q[\partial_t] &:= M = \frac{6\alpha}{\ell G(4-\alpha^2)(4-21\alpha^2)}\left[  2\sqrt{r_+ r_-}-\alpha(r_+ + r_-) \right]\,, \\
    Q[\partial_\theta] &:= J = \frac{6\sqrt{3}\,\alpha}{2\ell^2\mu^2G(1-\alpha^2)(4-\alpha^2)^{3/2}}\left[\alpha\sqrt{r_+r_-}(r_++r_-) -r_+r_-(\alpha^2+1) \right]\,,
\end{align}
respectively, where $\alpha$ and $\ell$ are related by Eq.~\eqref{lalphaWAdS}. In the limit $\alpha\to1$, the squashing effect disappears and the solution is locally equivalent to AdS$_3$. For this limiting case the solution above smoothly yields $\lambda_{(i)}=1/2$ for $ i=1,2,3$, and thus we have
\begin{align}
    \rho_0^2 = \frac{85\kappa e^2+\mu^2}{40\kappa e^2 \mu^2} \;\;\;\;\; \mbox{and} \;\;\;\;\; \frac{1}{\ell^2} = \frac{35\kappa e^2\mu^2(25\kappa e^2 + \mu^2)}{(85\kappa e^2 + \mu^2)^2}\,.
\end{align}
Another limit that is worth looking at is $e\to \infty$, where the solution above is seen to agree with the WAdS$_3$ black hole of NMG in vacuum, originally found in Ref.~\cite{Clement:2009gq}.

Notice that the black hole is $\eta$-Einstein with $\eta=\omega_3$. Thus, the NMG tensor satisfies a similar decomposition as Eq.~\eqref{etaEinstein}. This geometric property ultimately informs the energy-momentum tensor via the equations of motion. Hence, for arbitrary values of the parameters, the solution~\eqref{WAdSsol} yields a nonvanishing stress-energy tensor, $T_{\mu \nu }$, which can be written as a linear combination
\begin{equation}
T_{\mu\nu} = c_1\, g_{\mu\nu} + c_2\, K_{\mu\nu}\label{La34}
\end{equation}
where $K_{\mu \nu }$ is the tensor defined in Eq.~\eqref{Kmunu}, and where the coefficients are
\begin{subequations}
    \begin{align}\notag
    c_1 &= -\frac{1}{8\rho_0^4\alpha^2(\alpha^2-1)(21\alpha^2-4)}\Bigg[21\alpha^6(\lambda_{(3)}-\lambda_{(2)}^2)^2 + 16\lambda_{(2)}^2(\lambda_{(3)}-1)^2(1-5\alpha^2)  \\
    &\qquad + \alpha^4(63\lambda_{(3)}^2\lambda_{(2)}^2-20\lambda_{(2)}^4-86\lambda_{(3)}\lambda_{(2)}^2 -20\lambda_{(3)}^2 + 63\lambda_{(2)}^2  ) \Bigg]\,, \\
    c_2 &= \frac{4\left[\lambda_{(2)}^2(\lambda_{(3)}-1)^2 - \alpha^2(\lambda_{(3)}-\lambda_{(2)}^2)^2 \right] }{21\alpha^6 - 25\alpha^4+4\alpha^2} \,,
\end{align}
\end{subequations}
being finite as $\alpha \to 1$. Equation (\ref{La34}) implies that the YM field shifts the cosmological constant $\Lambda $ and the inverse of the mass parameter $\mu $. From the expressions for $c_1$ and $c_2$ one can verify that the WAdS$_3$ solution may exist even if the higher-curvature terms of NMG are not present, i.e. in the limit $\mu \to \infty $. That is to say, the YM terms suffice to support the black hole background in Einstein gravity, even when $\Lambda=0$. In some cases, the meron-type ansatz renders the YM configuration locally equivalent to an Abelian gauge field \cite{Canfora:2021nca}; in the case of the WAdS$_3$ black holes this seems to be different: while the WAdS$_3$ black holes are known to exist in the Einstein gravity coupled to the $U(1)$ gauge field theory \cite{Banados:2005da}, in that case, the CS term is needed to support the background. It is also worth mentioning that the solution we derived here persists when the gravitational CS term of Topologically Massive Gravity is also included. This merely induces a shift in the relation between the coupling constants of the theory and the parameters of the solution.

\section{Final Remarks\label{sec:V}}

In this work, we constructed exact warped-AdS black hole solutions in NMG minimally coupled to a non-Abelian gauge field. The solution asymptotes WAdS$_3$ at large radial distances and coincides with the NMG solution in vacuum found in \cite{Clement:2009gq} in the appropriate limit. This result might be relevant to the study of WAdS$_3$/WCFT$_2$ correspondence \cite{Anninos:2008fx, Detournay:2012pc,Donnay:2015iia,Donnay:2015vrb}, as rotating black hole solutions sourced by self-gravitating nonlinear matter fields are scarce. 

The theory under consideration is of fourth order and is non-linear both in the gravitational and the matter sector, which makes the problem of solving the field equations analytically quite intricate. Our strategy exploits the construction of $3$-dimensional spacetimes as Hopf vibrations of $2$-dimensional ones,  resulting in a simplification of the field equations. We considered a meron-type ansatz for the gauge field, which renders the problem tractable analytically. This ansatz turns out to be particularly useful to find exact solutions to this theory. Indeed, the solution can be perfectly embedded in Einstein-Yang-Mills theory as it is continuously connected to the latter when the higher-curvature terms are absent. 

Interesting questions remain open. For instance, in the AdS/CFT correspondence, holographic DC conductivities can be obtained from planar AdS black holes with Maxwell sources~\cite{Donos:2013eha,Andrade:2013gsa,Caldarelli:2016nni}. Holographic responses of Yang-Mills sources are worth exploring in chiral theories using the WAdS$_3$/WCFT$_2$ correspondence, as they could provide analytic expressions for the vacuum expectation value of strongly coupled non-Abelian currents. Additionally, null WAdS$_3$ spacetimes, also discussed in Ref.~\cite{Anninos:2008fx}, are known to be diffeomorphic to time-independent AdS waves observing the Schr\"odinger symmetry. Thus, looking for solutions of this type with self-gravitating Yang-Mills fields becomes relevant for describing condensed matter system holographically~\cite{Son:2008ye,Balasubramanian:2008dm}. Finally, it is known that squashed 3-spheres contribute nontrivially to the parity anomaly in three dimensions~\cite{Redlich:1983kn,Redlich:1983dv,Alvarez-Gaume:1984zst}. As this space bears a close resemblance to WAdS$_3$, a similar effect could be induced by the solution found here. We postpone a deeper study of these questions for future works. 

\begin{acknowledgments}
The authors thank Fabrizio Canfora and Rodrigo Olea for discussions. This work is supported by Agencia Nacional de Investigación y Desarrollo (ANID) through FONDECYT grants No 1240043, 1240048, 1230112, 1210500, 1221504, and 3220083.

\end{acknowledgments}

\appendix

\section{Other Solutions}

In order to further illustrate the usefulness of meron-type ans\"atze in this context, we provide explicit examples which extend the results presented above. We begin by returning to the squashed 3-sphere. Let us mention that the field ansatz \eqref{Aansatz1} is related to a geometric mapping of a three-sphere in spacetime into the three-sphere underlying the $SU(2)$ gauge group. Since the squashed three-sphere fibers over a two-sphere, a natural alternative is to consider a map from the sphere into the equator of the gauge group. This approach yields the following gauge potential
\begin{align}\notag
    A &= 2\lambda\,\Big[\left( \sin\varphi\,\rd \vartheta +\cos\varphi\cos\vartheta\sin\vartheta\,\rd \varphi\right)t_1
    \\ &\qquad - \left(\cos\varphi\,\rd \vartheta -\sin\varphi\cos\vartheta\sin\vartheta\,\rd \varphi\right)t_2
    - \sin^2\vartheta\,\rd \varphi\,t_3\Big]\,.\label{Aansatz2-b}
\end{align}
This choice, however, leads to a gauge field with a trivial topological charge, as the gauge connection depends only on two coordinates of the squashed 3-sphere. Inserting  Eq.~\eqref{ESquashed} and the ansatz~\eqref{Aansatz2-b} into the field equations~\eqref{eom}, we find that the latter reduces to a system of algebraic equations that admit the following solution
\begin{align}
    \lambda = \frac{1}{2} \; , \;\;\; \;\;\;\;\;
    \rho^2 = \frac{\alpha^2}{4k^2}\,,
\end{align}
with the coupling constants being related as follows 
\begin{align}
    \Lambda &= 
-\frac{k^{2} \left(21 \alpha^{6} e^{2} \kappa -93 \alpha^{4} e^{2} \kappa +88 \alpha^{2} e^{2} \kappa -42 \alpha^{2} k^{2}-16 e^{2} \kappa +40 k^{2}\right)}{2 e^{2} \kappa  \,\alpha^{2} \left(21 \alpha^{4}-25 \alpha^{2}+4\right)}
 \,, \\
    \mu^2 &= 
-\frac{k^{2} \kappa  \,e^{2} \left(21 \alpha^{4}-25 \alpha^{2}+4\right)}{2 \alpha^{4} e^{2} \kappa -2 \alpha^{2} e^{2} \kappa +4 k^{2}}
\,.
\end{align}
A real, Lorentzian version of the solution can also be obtained by performing the analytic continuation $\alpha\to i\alpha$. The relevance of solutions of this type is that they circumvent the need for anisotropy in the gauge field. 

A similar example we consider is the Nil geometry, which is one of the eight Thurston geometries that appear in connection to the geometrization conjecture. It can be associated to a Bianchi II space endowed with a transitive nilpotent group of diffeomorphisms. Locally, it remains invariant under the action of the Heisenberg group; see Refs.~\cite{Gegenberg:2003yz,Chernicoff:2018hpb,Flores-Alfonso:2021opl}. We write its Euclidean metric as
\begin{equation}
    \rd s^2 = \rho^2\left[\rd x^2+\rd y^2+(\rd z+x\rd y)^2\right]\,,
\end{equation}
where we denote $\rho$ as the radius. There are two, non-flat, left-invariant metrics on the Nil group with Lorentzian signature~\cite{Rahmani:1992}. One of them has a base space which is the Euclidean plane,
\begin{equation}\label{LorentzianNil1}
    \rd s^2 = \rho^2\left[\rd x^2+\rd y^2-(\rd t+x\rd y)^2\right],
\end{equation}
and the other is a fiber bundle over the Minkowski plane,
\begin{equation}\label{LorentzianNil2}
    \rd s^2 = \rho^2\left[-\rd t^2+\rd x^2+(\rd y+t\rd x)^2\right].
\end{equation}
They are sometimes referred to as warped flat spaces~\cite{Anninos:2009jt}. As pointed out in~Ref.~\cite{Mukhopadhyay:2013gja}, the first of these metrics admits the interpretation of a cross-section of the cylindrical Som-Raychauduri spacetimes~\cite{Som:1968}. On the other hand, metric~\eqref{LorentzianNil2} can be interpreted as a Kundt spacetime, a natural generalization of the $pp$-waves; see~\cite{Chow:2009vt,Boucetta:2022vny,Flores-Alfonso:2023fkd}. Using a geometrically similar ansatz for the gauge field as \eqref{Aansatz2-b}, we can show that NMG coupled to YM-CS theory also admits Nil manifolds as solutions. To show this, let us focus on the solution~\eqref{LorentzianNil1} and take the $SU(2)$ gauge field
\begin{align}\notag
    A &= 2\lambda\Bigg[\left(-\frac{\sin y\,\diff{x}}{\sqrt{1-x^2}}+x\cos y\sqrt{1-x^2}\,\diff{y}\right)t_1 \\ &\qquad + \left(\frac{\cos y\,\diff{x}}{\sqrt{1-x^2}}+x\sin y\sqrt{1-x^2}\,\diff{y}\right)t_2  - \left(1-x^2\right)\diff{y}\, t_3 \Bigg]\,.
\end{align}
Inserting this ansatz together with~\eqref{LorentzianNil1} into the field equations~\eqref{eom}, we find that the latter reduces to an algebraic system with the following solution
\begin{align}
    \lambda &= \frac{1}{2}\,, & \rho^2 &= \frac{1}{4k^2}\,, & \Lambda &= \frac{k^2\left(\kappa e^2+2k^2\right)}{2\kappa e^2}\,, & \mu^2 &= \frac{21\kappa e^2 k^2}{2\left(\kappa e^2 - 2k^2 \right)}\,,
\end{align}
which holds provided $\kappa e^2-2k^2\neq0$. In fact, if the latter condition is not satisfied, there are no solutions for $\mu\neq0$. Notice also that this solution does not exist in the absence of the CS term. A similar solution for Eq.~\eqref{LorentzianNil2} exists, which is also valid for $\kappa e^2-2k^2=0$.

Other interesting solutions of NMG that support a non-trivial gauge field configuration are AdS-waves, which are analogs of $pp$-waves but in AdS$_3$, cf.~\cite{Ayon-Beato:2005gdo,Ayon-Beato:2009cgh}. The metric of these solutions takes the form
\begin{align}\label{AdSwaves}
    \diff{s^2} = \frac{\rho^2}{y^2}\left[-2\diff{u}\diff{v}-F(u,y)\diff{u^2}+\diff{y^2}\right]\,,
\end{align}
where $u\in \mathbb{R}$ and $v\in \mathbb{R}$ are null coordinates, $\rho\in \mathbb{R}_+$ is a constant, and $F(u,y)$ is a function that represents the profile of the wave and is obtained by solving the field equations. Locally AdS$_3$ corresponds to the particular case $F=\text{const}$. In these coordinates, the $SU(1,1)$ gauge field $A = \lambda \omega_i\eta_i$ is parameterized as
\begin{align}
    A = &2\lambda\Bigg[  \left( \frac{v^2-2v+2}{y^2}\rd u +\frac{\rd v}{2}-  \frac{v-1}{y}\rd y
    \right)\eta_1\notag + \left( \frac{v(v-2)}{y^2}\rd u +\frac{\rd v}{2}-  \frac{v-1}{y}\rd y
    \right)\eta_2\notag\\
    &\quad +\left(\frac{2(v-1)}{y^2}\rd u -  \frac{1}{y}\rd y
    \right)\eta_3\Bigg]\,.
\end{align}
Inserting these ans\"atze into the YM-CS equations~\eqref{eomA}, we find that the matter equations are solved if
\begin{align}
   \lambda = \frac{1}{2} - \frac{k\rho}{2} \;\;\;\;\; \mbox{and} \;\;\;\;\; F(u,y) = c(u)\,y^{1-k\rho}\,, 
\end{align}
where $c(u)$ is an arbitrary function of the null coordinate $u$. Notice that we have not yet considered at this point the NMG equations. In the vacuum case, the equations allow for two types of solutions in accordance with the degrees of freedom of the theory \cite{Ayon-Beato:2009cgh}. However, in the present case the matter equations restrict the system as above. In particular, notice that the system does not allow for logarithmic modes and so the Breitenlohner-Freedman bound cannot be saturated \cite{Breitenlohner:1982bm}. Interestingly enough, the YM-CS equations do allow for Schr\"odinger symmetry to arise \cite{Duval:2008jg}. In the equation above this occurs for the special value $k\rho=3$. Lastly, inserting these solutions into the remaining field equations, we find that they are solved by
\begin{subequations}
\begin{align}
     \rho^2 &= \frac{\kappa e^2 - 2\mu^2}{2\left[\kappa e^2\left(k^2-\mu^2 \right) - k^2\mu^2 \right]} \,, \\
     \Lambda &= -\frac{4\mu^6\left(5\kappa e^2+4k^2\right) + 4\mu^4\left(k^4 -5\kappa e^2 k^2 - \kappa^2 e^4  \right) - 7\mu^2\kappa e^2 k^4 + 4\kappa^2 e^4 k^4}{4\mu^2\left(\kappa e^2-2\mu^2 \right)^2}\,.
\end{align}    
\end{subequations}

This manifestly shows that the strategy followed throughout this paper turns out to be useful for finding exact, analytic solutions of this type of non-linear, higher-derivative theories.

\bibliography{References}

\end{document}